\documentclass[sigconf,natbib=false]{acmart}
\usepackage{cite}
\usepackage{multirow}
\usepackage{colortbl}

\AtBeginDocument{%
  }

\renewcommand\footnotetextcopyrightpermission[1]{}
\renewcommand{\shortauthors}{}

\begin{document}

\title{ScaleLUT: A Fully-Parallel Configurable LUT-Based Accelerator for Real-Time Multi-Scale Super-Resolution}

% ===== Compact author block =====

\author{
\fontsize{9.5pt}{11pt}\selectfont
Boyu Li$^{1,*}$,
Chenchen Ding$^{1,*}$,
Zhilin Ai$^{1}$,
Wenqing Shi$^{1}$,
Baizhou Jiang$^{1}$,
Wenyong Zhou$^{1}$, \\
Binxiao Huang$^{1}$,
Jiachen Ren$^{1}$,
Hao Yu$^{2}$,
and Ngai Wong$^{1,\dagger}$
\\[3pt]
$^{1}$The University of Hong Kong (HKU), Hong Kong
\\
$^{2}$Southern University of Science and Technology, Shenzhen, China
}

\thanks{$*$ Equal contribution.
This project is supported in part by the Theme-based Research Scheme (TRS)
projects T45-701/22-R and T41-517/25-N, and GRF Project 17203224 of the
Research Grants Council (RGC), Hong Kong SAR; in part by the AVNET-HKU
Emerging Microelectronics \& Ubiquitous Systems (EMUS) Lab; and in part by
the Shenzhen Science and Technology Program under Grant Nos.
KQTD20200820113051096, JCYJ20220818100217038, and ZDCY20250901101004005.
$\dagger$ Corresponding author: Ngai Wong
(e-mail: nwong@eee.hku.hk).
}

% \affiliation{%


\begin{abstract}

Real-time super-resolution (SR) remains challenging for edge devices. Although deep-learning-based SR methods achieve high reconstruction quality, their heavy multiply-accumulate (MAC) operations incur substantial resource and power costs, limiting practical deployment. Lookup-table (LUT)-based SR offers a more efficient alternative by replacing convolutional inference with precomputed table queries. However, existing LUT-based methods still suffer from limited speed, high storage overhead, and restricted scalability across upsampling factors. This paper presents ScaleLUT, a hardware-oriented LUT design framework and fully parallel reconfigurable accelerator for real-time multi-scale SR. By jointly optimizing LUT design and hardware implementation, ScaleLUT achieves competitive SR performance with high hardware efficiency. Specifically, a hardware-friendly YUV-domain strategy and a power-of-two kernel design with rotation ensemble are introduced to improve receptive-field (RF) coverage while reducing LUT dimensionality, where the latter further simplifies hardware by replacing division operations with shift operations. Together, these designs yield an 18.4\% memory reduction over state-of-the-art (SOTA) LUT-based SR methods. Furthermore, ScaleLUT supports arbitrary input resolutions and configurable $\times 2^n$ upsampling factors with a deeply pipelined and massively parallel architecture that fully exploits the multiplication-free nature of LUT inference. Implemented on a Xilinx ZCU102 FPGA, ScaleLUT achieves real-time 4K SR at 95.3 FPS for $\times2$ upscaling at 300 MHz. Compared with existing SR accelerators, it uses at least 58.6\% fewer LUTs, 41.1\% fewer flip-flops, zero DSPs, and 42.0\% lower power. It also delivers 10$\times$ and 1.2$\times$ speedups over the best CPU-based SR implementation and prior FPGA-based SR accelerators, respectively. Overall, ScaleLUT demonstrates the effectiveness of joint LUT algorithm–hardware co-design for practical and energy-efficient edge SR deployment.
\vspace{-0.25cm}
\end{abstract}


\maketitle
\pagestyle{empty}
\thispagestyle{empty}

%%%%%%%%%%%%%%%%%%%%%%%%%%
%       Introduction
%%%%%%%%%%%%%%%%%%%%%%%%%%
\vspace{-0.25cm}
\section{Introduction}
\label{sec:intro}

Image super-resolution (SR) reconstructs high-resolution (HR) images from low-resolution (LR) inputs, and is widely used in 4K display enhancement and video streaming. Classical SR methods, including interpolation~\cite{Bevilacqua2012,Kirkland2010,1981keys,Nuno2005} and sparse-coding approaches~\cite{radu2013,radu2015}, are computationally efficient but struggle to recover high-frequency details. Deep-learning-based SR (DLSR) models~\cite{SRCNN,EDSR,RCAN,ESPCN,ahn2018,wang2019,IMDN,FMEN,taito2016,he2018fpga,kim2018real,chang2020,sun2021fpga,shi2019winograd} greatly improve reconstruction quality, but their convolution-heavy computation incurs substantial MAC operations, latency, and power consumption, limiting real-time deployment on edge platforms.

\begin{figure}[t] \setlength{\belowcaptionskip}{-0.0cm}  \includegraphics[width=0.74\columnwidth]{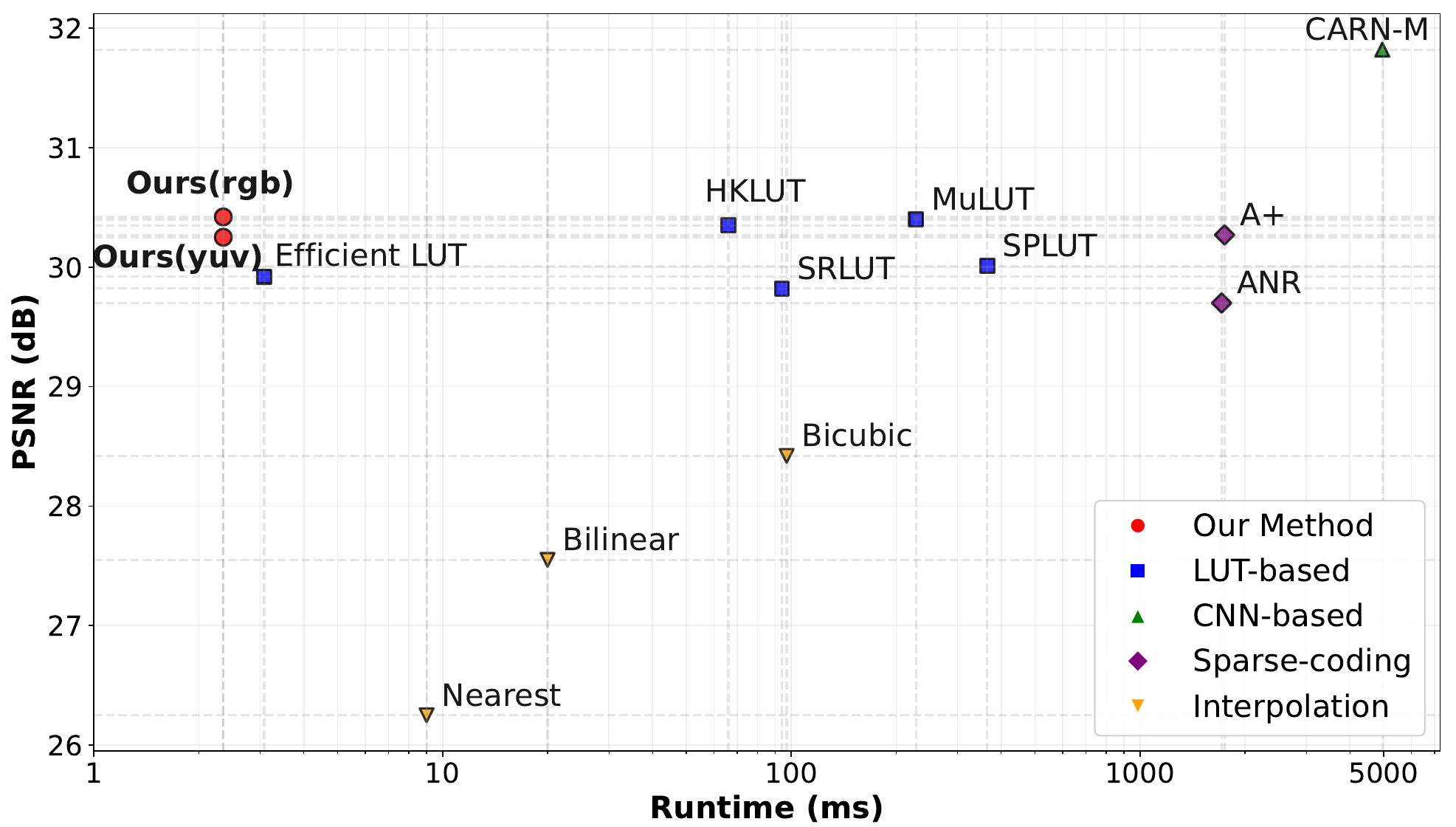} \caption{ Comparison of PSNR on Set5 for $4\times$ SR versus runtime (1280$\times$720 output). Interpolation methods (inverted triangles), LUT-based methods (squares), sparse-coding methods (diamonds), and DLSR methods (triangles) are included. Our method achieves comparable or better PSNR than prior LUT-based approaches while running faster.
} 
   \vspace{-0.6cm}
 \label{fig: compare_psnr_runtime} \end{figure}

Recent LUT-based SR methods~\cite{Jo2021SRLUT,Li2022MuLUT,Ma2022SPLUT,Liu2023RCLUT,huang2024HKLUT,libdlut_sid,li2025bdlut_jsid,li2024,Li2026edgelut,ai2026hybridlut} replace convolutional inference with precomputed table queries, largely eliminating MAC-heavy processing. Existing studies have explored single-LUT inference~\cite{Jo2021SRLUT}, multiple cooperating LUTs~\cite{Li2022MuLUT}, series-parallel LUT structures~\cite{Ma2022SPLUT}, reconstructed-convolution-based LUTs~\cite{Liu2023RCLUT}, compact LUT models~\cite{huang2024HKLUT}, and storage-efficient variants~\cite{li2024}. However, these methods still face a fundamental trade-off among receptive-field (RF) coverage, LUT size, memory access cost, and deployment efficiency. Enlarging the RF usually increases LUT complexity and storage demand, while independent RGB-channel processing either lowers throughput or replicates LUTs, easily exceeding SRAM budgets on FPGAs and ASICs. Therefore, existing LUT-based SR methods remain largely algorithm-centric and are not well aligned with hardware execution pipelines.

To address these limitations, we propose ScaleLUT, a hardware-oriented LUT design framework for real-time multi-scale SR on resource-constrained edge platforms. As shown in Fig.~\ref{fig: compare_psnr_runtime}, ScaleLUT achieves comparable or better PSNR than prior LUT-based SR methods with significantly lower runtime. Unlike previous methods that mainly optimize algorithm-level efficiency, ScaleLUT jointly considers LUT construction and hardware execution through systematic design-space exploration, as illustrated in Fig.~\ref{fig:NAS}. Based on this co-design strategy, ScaleLUT provides a fully parallel and configurable LUT-based accelerator for real-time high-resolution SR. We implement the complete system on a Xilinx ZCU102 FPGA and validate its practicality in real-time video SR scenarios. The main contributions are summarized as follows:

\begin{itemize}
  % \item We adopt a hardware-friendly kernel design with power-of-two kernel counts and a rotation-ensemble scheme that guarantees full receptive-field coverage, thereby improving both SR performance and hardware efficiency. Extensive evaluations show that both the RGB and YUV variants of ScaleLUT achieve competitive PSNR, while the YUV model reduces on-chip memory usage by two-thirds compared with the RGB version and by at least 18.4\% over SOTA LUT-based methods.
  % \item We propose a novel hardware-oriented LUT search framework that jointly optimizes SR performance and hardware efficiency, achieving an improved balance between reconstruction quality and implementation cost. Through systematic search and ablation analysis, the framework identifies an optimal combination of power-of-two kernel designs, which replaces division operations with shift operations, significantly simplifying hardware complexity and improving SR performance. In addition, a YUV-domain formulation reduces on-chip memory usage by two-thirds compared to the RGB-domain formulation and by at least 18.4\% over SOTA LUT-based methods with negligible accuracy loss, further enhancing hardware practicality. Moreover, a bit decomposition strategy is introduced to replace interpolation, enabling a zero DSP LUT-based SR implementation on FPGA for the first time.

  \item We propose a novel hardware-oriented LUT design framework with systematic design space exploration that jointly optimizes SR performance and hardware efficiency, achieving an improved balance between reconstruction quality and implementation cost. Through search and ablation analysis, the framework identifies an optimal combination of power-of-two kernel designs, which replaces division operations with shift operations, significantly simplifying hardware complexity and improving SR performance. In addition, a YUV-domain formulation reduces on-chip memory usage by two-thirds compared to the RGB-domain formulation and by at least 18.4\% over SOTA LUT-based methods with negligible accuracy loss, further enhancing hardware practicality. Moreover, a bit decomposition strategy is introduced to replace interpolation, enabling a zero DSP LUT-based SR implementation on FPGA for the first time.

  \item We are the first configurable LUT-based SR accelerator supporting arbitrary input resolutions and any $\times 2^n$ upsampling factors. The architecture leverages deep pipelining and massive parallelism to fully exploit the multiplication-free nature of LUT inference, enabling wide adaptability to practical edge-side deployment scenarios.

  \item We demonstrate superior resource, energy, and performance efficiency. On ZCU102 FPGA, the complete accelerator and HDMI subsystem achieve real-time 4K video SR with at least 58.6\% fewer LUTs, over 41.1\% fewer flip-flops, zero DSP usage, and at least 42\% lower power consumption compared with existing SR accelerators. Moreover, our design achieves a $10\times$ speedup over CPU-based SR algorithms and a $1.2\times$ speedup over the best FPGA-based accelerators.
  \vspace{-0.15cm}
\end{itemize}

%%%%%%%%%%%%%%%%%%%%%%%%%%
%       Related Work
%%%%%%%%%%%%%%%%%%%%%%%%%%

\section{Background and Related Work}

\subsection{Super-Resolution Algorithms}

Classical SR methods, including interpolation~\cite{Bevilacqua2012,Kirkland2010,1981keys} and sparse-coding approaches~\cite{radu2013,radu2015}, offer low computational cost but suffer from limited reconstruction fidelity or irregular computation patterns, making them insufficient for high-quality real-time SR deployment. Deep-learning-based SR methods~\cite{SRCNN,EDSR,RCAN,ESPCN,ahn2018,IMDN,FMEN,wang2019} significantly improve reconstruction quality by learning nonlinear LR-to-HR mappings. However, their convolution-heavy architectures require substantial MAC operations, feature-map storage, and memory bandwidth, leading to high latency and power consumption on resource-constrained edge devices.

\begingroup
\emergencystretch=.75em
LUT-based SR methods~\cite{Jo2021SRLUT,Li2022MuLUT,Ma2022SPLUT,Liu2023RCLUT,huang2024HKLUT,libdlut_sid,li2025bdlut_jsid,li2024,Li2026edgelut,ai2026hybridlut} replace convolutional inference with precomputed table queries, providing a promising alternative for efficient SR. Representative works have improved LUT-based inference through single-LUT transfer~\cite{Jo2021SRLUT}, multiple cooperating LUTs~\cite{Li2022MuLUT}, series-parallel structures~\cite{Ma2022SPLUT}, reconstructed-convolution-based LUTs~\cite{Liu2023RCLUT}, compact LUT models~\cite{huang2024HKLUT}, and storage-efficient designs~\cite{li2024}. Nevertheless, most existing LUT-based SR methods remain algorithm-centric. Their LUT organization, receptive-field coverage, memory footprint, and execution patterns are not jointly optimized with hardware constraints, leaving a gap between LUT-level efficiency and real-time high-resolution deployment.
\par
\endgroup

\subsection{Hardware Accelerators for Super-Resolution}

Early FPGA-based image scaling accelerators, such as bicubic interpolation designs~\cite{Nuno2005}, achieve low hardware cost and simple streaming dataflow but provide limited visual quality. CNN-based SR accelerators on FPGAs or ASICs~\cite{taito2016,chang2020,kim2018real,sun2021fpga,he2018fpga,shi2019winograd} map deep SR networks onto systolic arrays, line buffers, or customized datapaths. However, their large numbers of convolutional layers, high-dimensional feature maps, and MAC-intensive computation result in considerable LUT/FF usage, on-chip memory demand, and dynamic power. Quantized FPGA deployment further requires customized processing elements~\cite{float_int_FPGA}, increasing design complexity and resource pressure.

Dedicated LUT-based SR hardware remains underexplored. Although LUT inference naturally avoids MAC-heavy computation, existing LUT-oriented designs still face challenges in jointly balancing receptive-field coverage, sampling effectiveness, arithmetic simplicity, memory size, and scalability across upsampling factors. Efficient-LUT~\cite{li2024} shows the potential of LUT-based hardware acceleration, but a systematic algorithm--hardware co-design framework for real-time, memory-efficient, and configurable multi-scale SR is still lacking.

%%%%%%%%%%%%%%%%%%%%%%%%%%
%       Methodology
%%%%%%%%%%%%%%%%%%%%%%%%%%

\section{LUT-based Neural Network Generation}
\label{sec:algo}

\subsection{Framework Design and Design Space Exploration}
\label{subsec:framework_design}

\begin{figure}[!t]
\centering
\includegraphics[width=0.75\columnwidth]{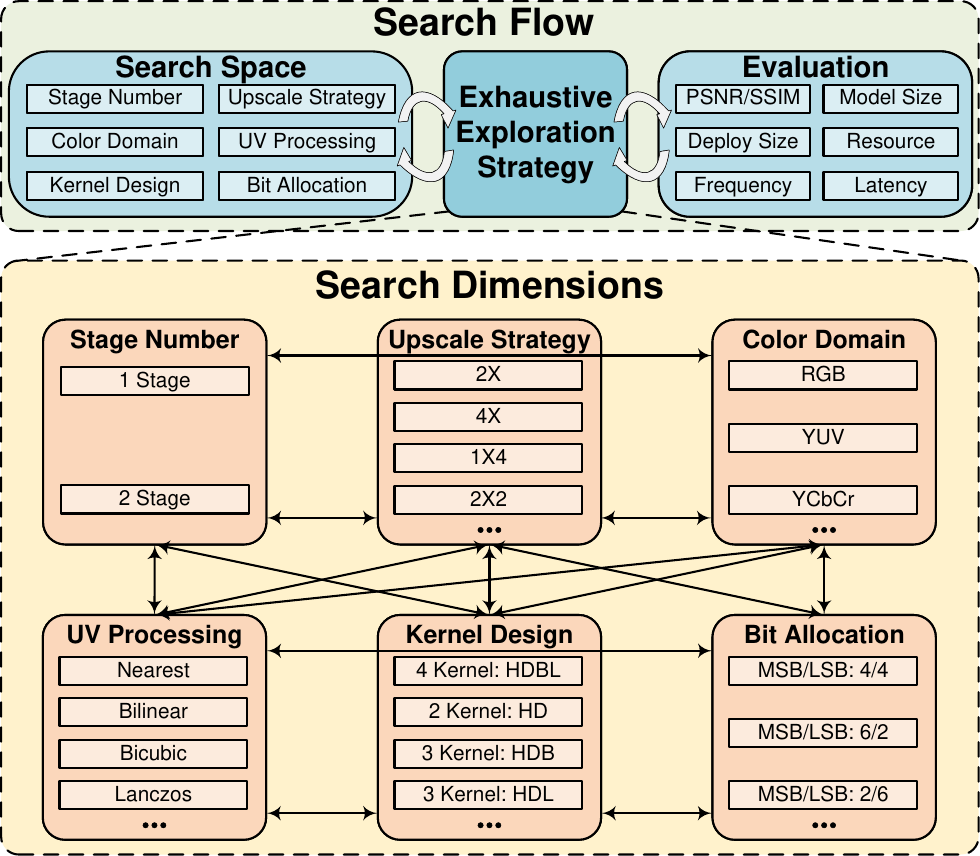}
\caption{Overview of the proposed hardware-oriented LUT design framework with exhaustive design space exploration.}
\vspace{-0.6cm}%%减小图片上间隔
\label{fig:NAS}
\end{figure}

As shown in Fig.~\ref{fig:NAS}, ScaleLUT is derived from a hardware-oriented design framework instead of a manually fixed LUT structure. The framework jointly explores reconstruction quality and implementation efficiency, enabling the final design to balance SR performance with real-time hardware deployment cost.

The search space covers stage number, upscale strategy, color domain, UV processing, kernel design, and bit allocation. Each candidate is evaluated using both algorithmic metrics, including PSNR/SSIM and model size, and hardware metrics, including deploy memory, resource usage, frequency, and latency. These dimensions directly affect hardware cost: the color domain determines the number of LUT replicas, UV processing affects chrominance-path complexity, kernel design changes adder-tree depth and normalization logic, and bit allocation determines LUT address width and RAM size. With hardware-level cost modeling of dividers, multipliers, shifters, adders, and memories, the proposed exhaustive exploration identifies the design point with the best quality--efficiency trade-off.

\begin{figure}[t]
\centering
\includegraphics[width=0.75\columnwidth]{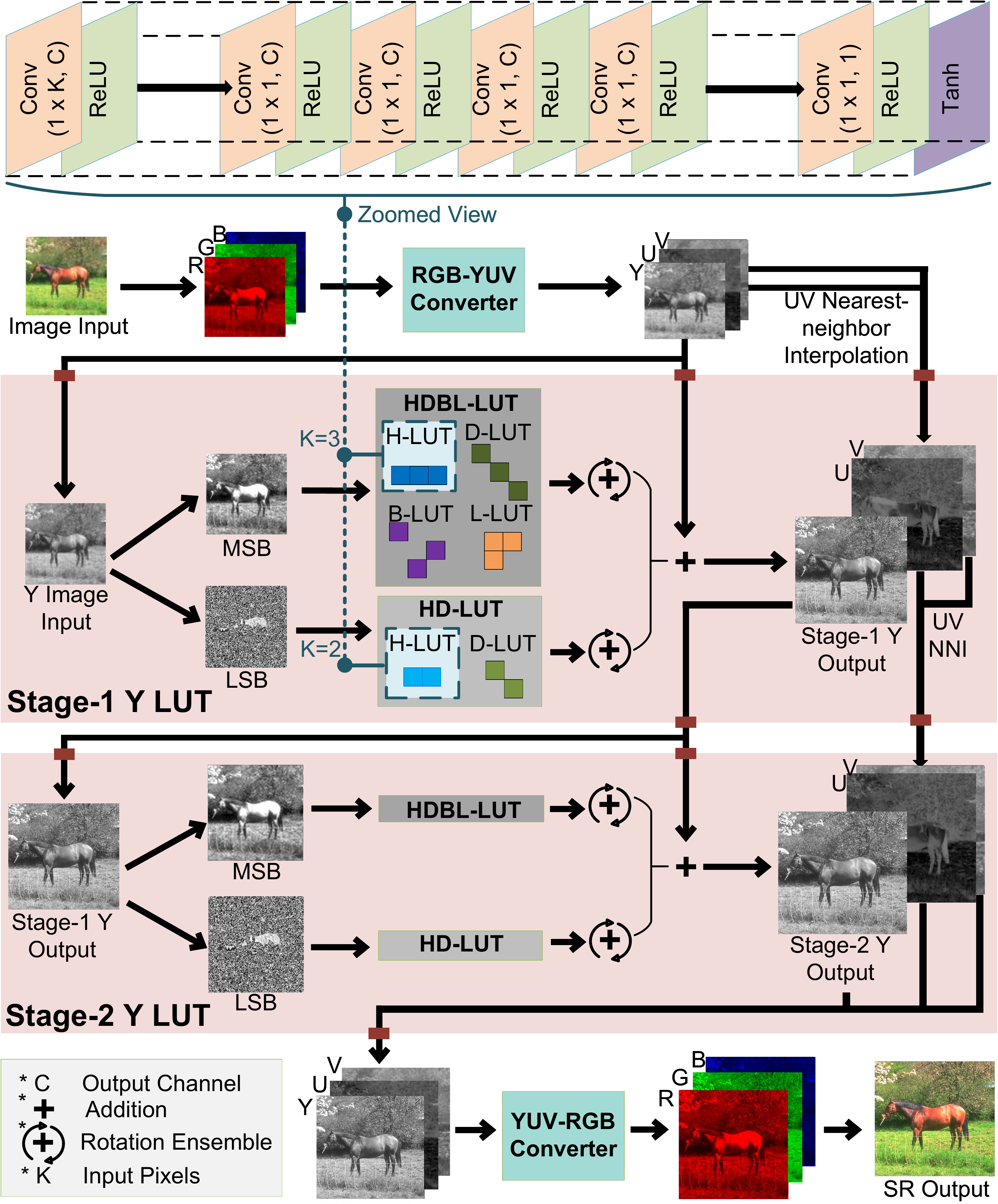}
\caption{Overall architecture of LUT and corresponding network. An efficient SR system employing a two-stage dual-branch LUT architecture with rotation ensemble for the Y channel and interpolation for the UV channels.}
\vspace{-0.6cm}%%减小图片上间隔
\label{fig:lut_net}
\end{figure}

\subsection{LUT Network Architecture}
\label{subsec:lut_network}

\begingroup
\emergencystretch=.5em
Based on the selected design point, we construct a hardware-efficient LUT network for YUV-domain SR, as shown in Fig.~\ref{fig:lut_net}. Since YUV is widely used in practical imaging and video systems, ScaleLUT applies LUT-based SR only to the luminance channel Y, while the chrominance channels U and V are separately upscaled and merged back after luminance restoration. Compared with RGB-domain LUT inference on three channels, this strategy significantly reduces LUT storage and on-chip memory usage while maintaining competitive reconstruction quality.
\par
\endgroup

The proposed network adopts a two-stage dual-branch architecture for the Y channel. The 8-bit luminance input is decomposed into 4-bit most significant bits (MSBs) and 4-bit least significant bits (LSBs), which are processed by two asymmetric branches. The MSB branch captures dominant structural information, while the LSB branch refines local details. Their outputs are fused with the center pixel to generate the first-stage result, and the same structure is reused in the second stage for further refinement.

\begin{figure}[!t]
\centering
\includegraphics[width=0.62\columnwidth]{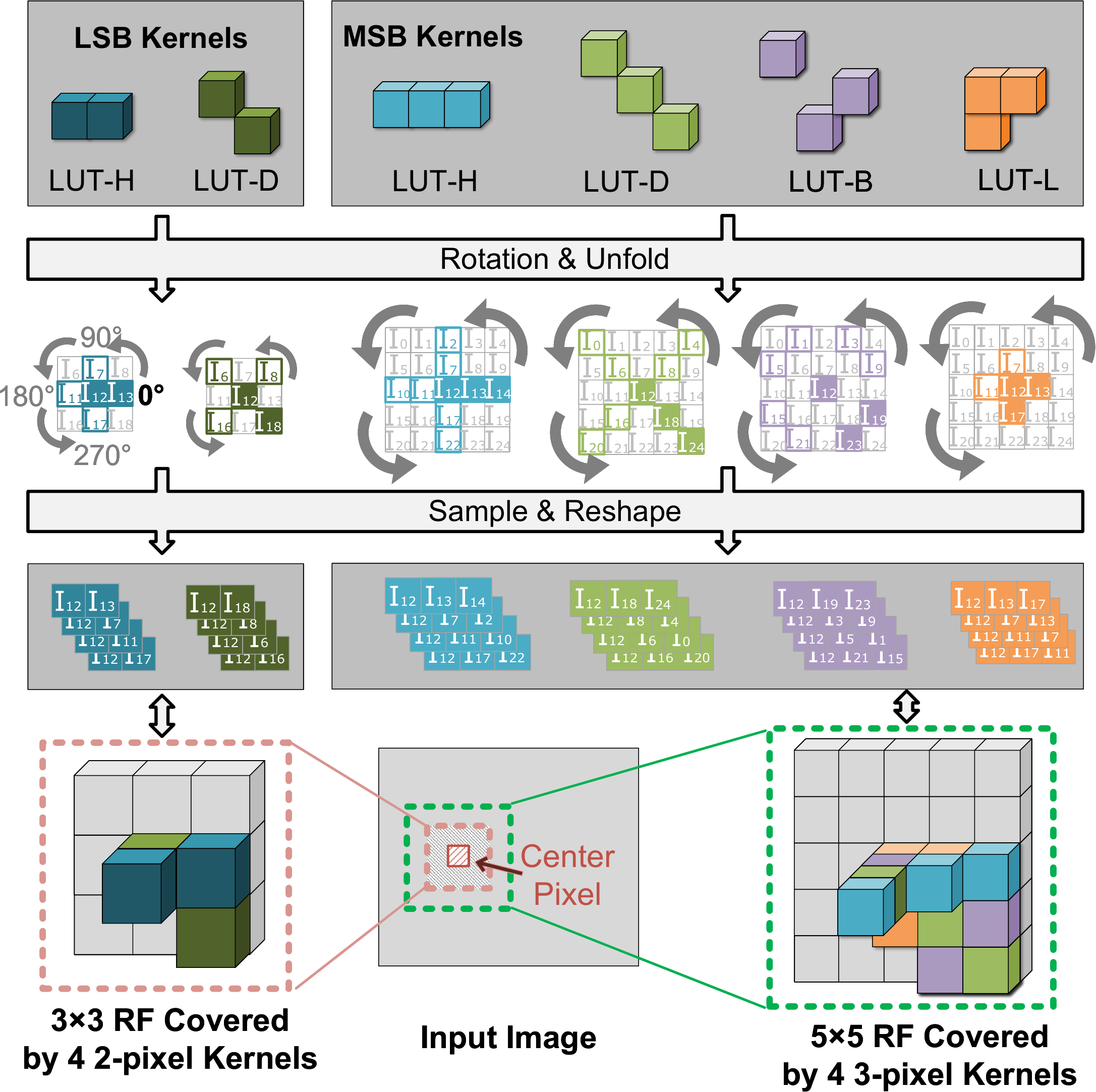}
\caption{Power of two kernels with rotation ensemble: 3-pixel kernels HDBL for MSB branch, 2-pixel kernels HD for LSB branch.}
\vspace{-0.4cm}%%减小图片上间隔
\label{fig:lut_kernel}
\end{figure}

To balance receptive-field coverage and storage cost, ScaleLUT employs asymmetric power-of-two kernels with rotation ensemble, as shown in Fig.~\ref{fig:lut_kernel}. The MSB branch uses four 3-pixel kernels, H/D/B/L, whose rotations cover a $5\times5$ RF without high-dimensional LUTs. The LSB branch adopts two 2-pixel kernels, H/D, to cover a compact $3\times3$ RF. This design allocates more representation capacity to the structurally important MSB branch while keeping the LSB branch compact, achieving a better balance between reconstruction capability and storage efficiency.

\label{sec:hardware}
\begin{figure*}[t]
    \centering
    \includegraphics[width=0.59\linewidth]{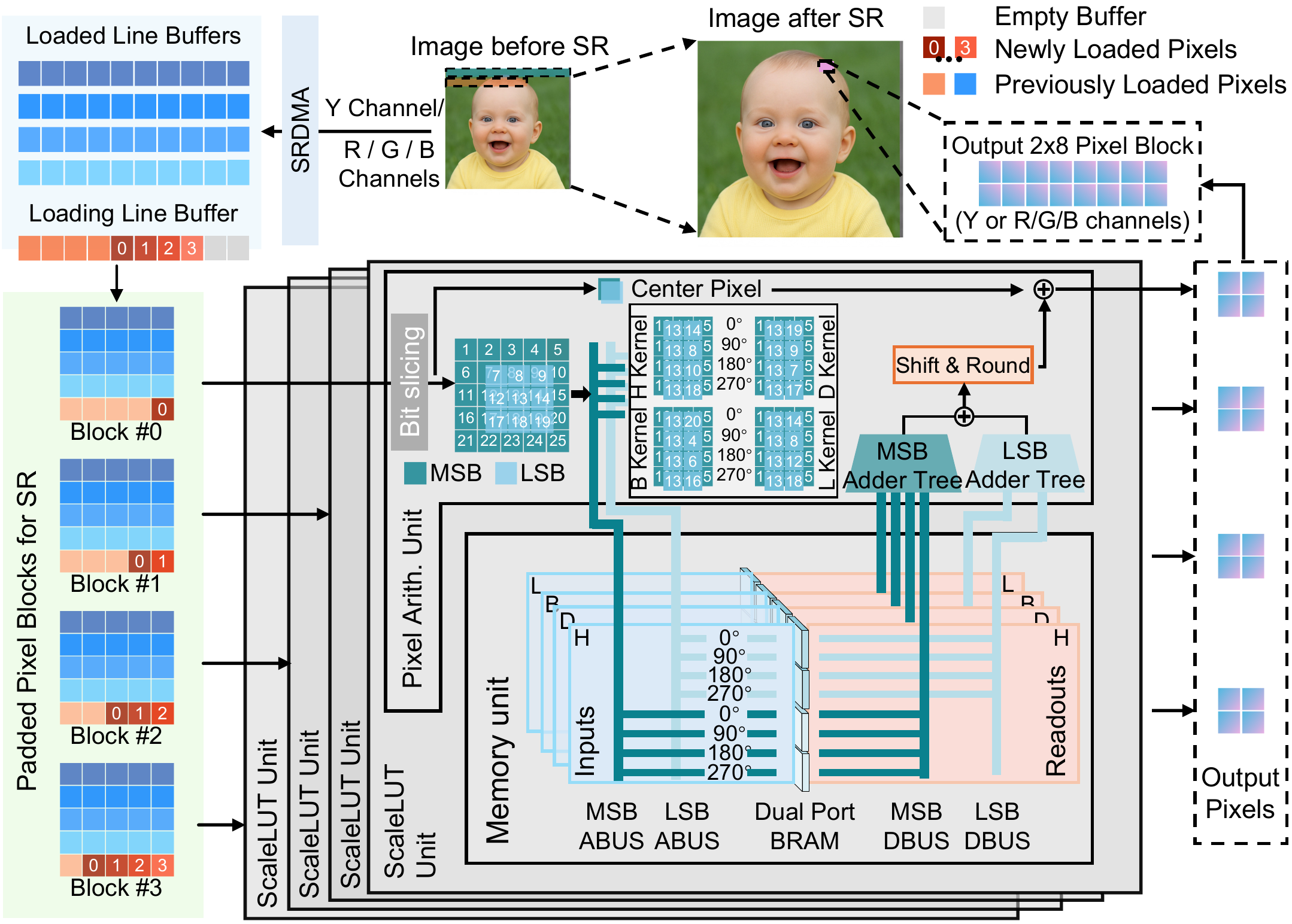}
    \caption{Architecture of LUT-based SR accelerator for processing Y channel in YUV format images or R/G/B channels in RGB images. The elements include DMA and line buffers for efficient pixel padding, 4 ScaleLUT units with
 dual-port BRAM access for high-throughput pixel block computation.}
 \vspace{-0.5cm}%%减小图片上间隔
    \label{fig:arch_unit}
\end{figure*}

% \vspace{-0.1cm}
\section{Proposed LUT-based Configurable SR Accelerator Architecture}

\subsection{Architecture of LUT-based SR accelerator}
To support real-time processing of high-resolution video streams, the proposed LUT-based configurable SR accelerator adopts a fully pipelined and parallel architecture, as shown in Figure \ref{fig:arch_unit}. Input frames are first fetched from the DDR4 memory via the SRDMA engine and buffered on chip by five line buffers. The image data are streamed into these line buffers row by row in a circular fashion. During operation, four line buffers (loaded line buffers) hold previously received rows of pixels, while the remaining buffer (loading line buffer) stores the most recently loaded row.

Whenever a new 1$\times$4 group of pixels arrives in the loading line buffer, it is combined with pixels in the loaded line buffers to construct four 5$\times$5 pixel blocks, which are dispatched in parallel to four ScaleLUT units. Each ScaleLUT unit consists of a Pixel Arithmetic Unit and a Memory Unit. As introduced in Section~\ref{sec:algo}, the Pixel Arithmetic Unit slices the 5$\times$5 block into MSB and LSB components according to the predefined 5$\times$5 and 3$\times$3 RFs, and generates four address kernels (H/D/B/L) that are forwarded to the Memory Unit through the address bus (ABUS). The Memory Unit organizes LUT entries into 16 independent banks, with MSB and LSB tables each mapped to eight banks. In the read phase, the MSB entries for the H/D/B/L kernels and the LSB entries for the H/D kernels are accessed in parallel. Retrieved data are returned through the data buses and fed into dedicated adder trees, which perform accumulation and rounding to generate intermediate pixel values. Each intermediate value is then added to the center pixel of the corresponding input block, producing the 2$\times$2 super-resolved outputs.
With this organization, each ScaleLUT unit produces one 2$\times$2 block per cycle, and the four units together generate a 2$\times$8 output block, which is written back to DDR4 memory through the AXI bus to complete one iteration of the SR pipeline.

For different color formats, we implement two SR accelerator variants. RGB accelerator processes all three channels (R, G, and B) through ScaleLUT units in parallel, and the super-resolved channels are concatenated to form the output image. YUV accelerator contains RGB2YUV and YUV2RGB converters for image input and output. For images converted to YUV, only Y channel is processed by ScaleLUT units, while the upscaled U and V channels are obtained by replicating the UV values at the center pixel of each 5$\times$5 block, which are then concatenated with the super-resolved Y channel to form the output frame.

\begin{figure}[!t]
    \centering
    \includegraphics[width=0.78\linewidth]{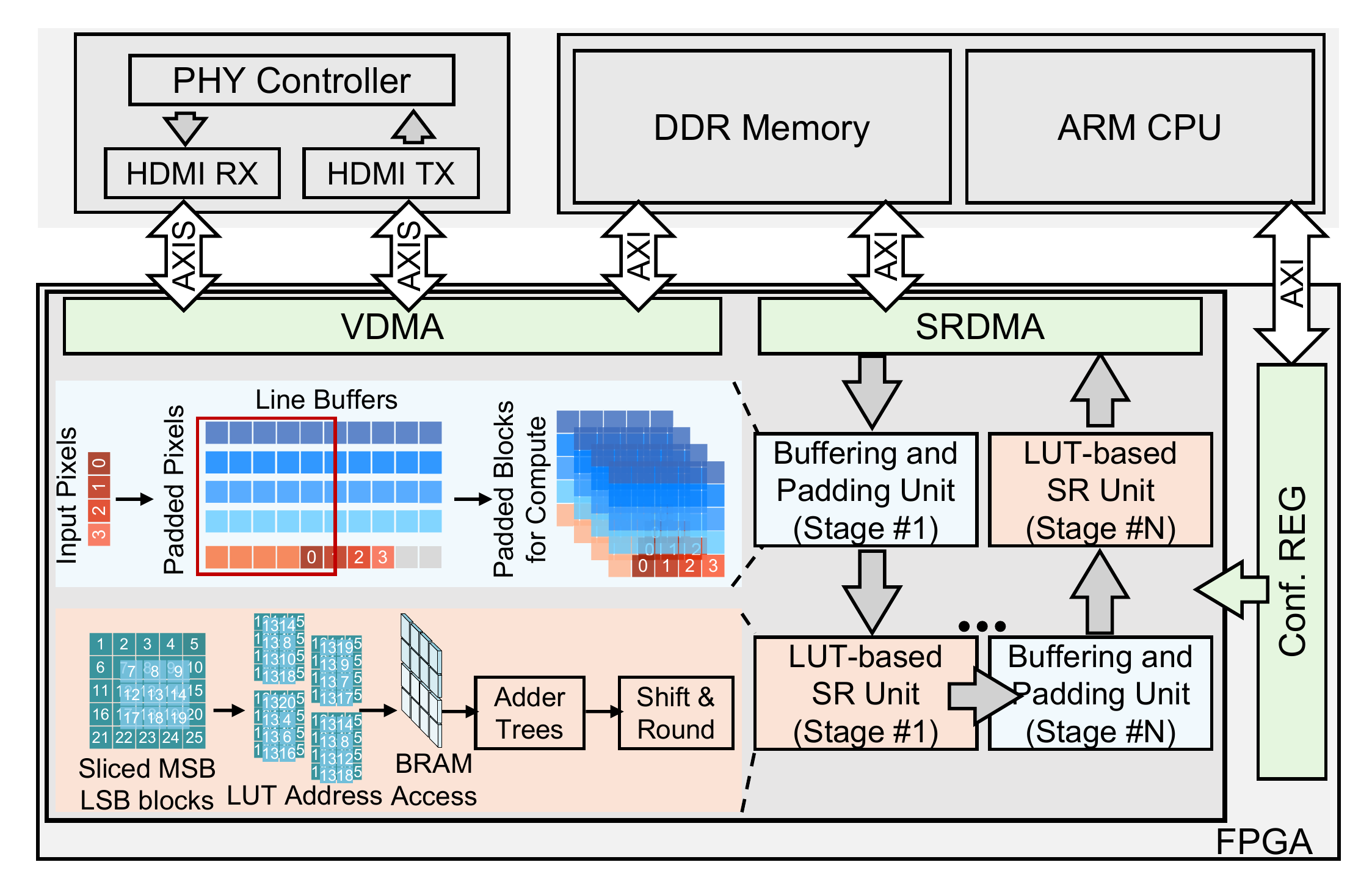}
    % \vspace{-1em}
    \caption{Overall architecture of the LUT-based SR system implemented with ZCU102 FPGA platform}
    \vspace{-0.5cm}%%减小图片上间隔
    \label{fig:arch_deploy_1}
\end{figure}

\begin{figure}[!t]
    \centering
    \includegraphics[width=0.78\linewidth]{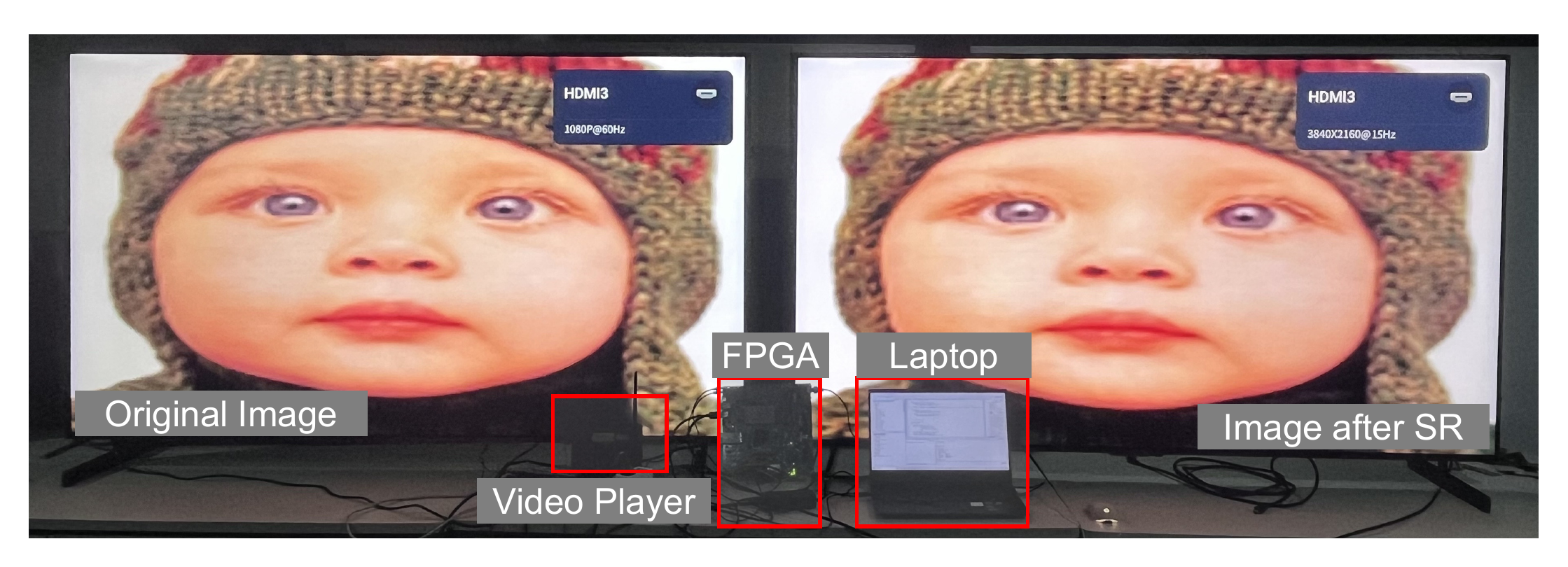}
    \caption{Comparison of 1080p input image and 4K output image after super-resolution in the deployed system.}
    \vspace{-0.6cm}%%减小图片上间隔
    \label{fig:arch_deploy}
\end{figure}

\subsection{Configurable Deployment System for Real-time SR Applications}
To support practical deployment in real-time display and configurable SR applications, we further develop a video streaming system over HDMI as shown in Figure \ref{fig:arch_deploy_1}. 
Video is streamed through the HDMI PHY interface. The HDMI RX subsystem captures raw input from an external player, converts it into DE‑Sync Video, and sends it to the VDMA via the AXI‑Stream (AXIS) bus, which then writes the data to DDR memory on the FPGA board through the AXI bus. After the SRDMA writes the processed results back to DDR memory, the HDMI TX subsystem retrieves the data through the VDMA and outputs the enhanced video to a 4K display.

To enable more flexible operation, we introduce a process management block that receives external configuration signals from the CPU via the AXI bus. This module is integrated above the LUT‑based SR datapath and provides a compact set of configuration registers to flexibly control the input and output resolutions as well as the number of SR stages. Specifically, it supports three SR modes—2$\times$ SR, 4$\times$ SR, and 8$\times$ SR—by processing images with 1, 2, and 3 stages, respectively, without introducing additional hardware overhead.

Figure~\ref{fig:arch_deploy} presents a photograph of the SR system based on ScaleLUT. The left screen displays the original 1080p low-resolution image, while the right screen shows the 4K SR output generated by the ScaleLUT algorithm. Note that due to the adaptive scaling chip integrated in the TV, its onboard upscaling algorithm enlarges the 1080p low-resolution image to fill the entire 4K display panel, but the performance of this built-in upscaling algorithm is inferior to that of the 4K output processed by ScaleLUT.

%%%%%%%%%%%%%%%%%%%%%

\begin{table*}[t]
\renewcommand{\arraystretch}{0.7}
\caption{Quantitative comparison (PSNR/SSIM) and model size for 4$\times$ super-resolution, where all methods upscale images to 1280$\times$720 across five benchmark datasets. (SC: Sparse-coding)}
\vspace{-0.3cm}
% \vspace{-0.1cm}
\centering{
\setlength{\tabcolsep}{0.1mm}{
\begin{tabular}{lcccccccccc}
\toprule
\multirow{2}{*}{\textbf{Cat.}} & \multirow{2}{*}{\textbf{Method}} & \multirow{2}{*}{\textbf{Platform}}& \multirow{2}{*}{\textbf{Runtime}}& \multirow{2}{*}{\textbf{Size}}& \textbf{Actual} & \textbf{Set5}  & \textbf{Set14} & \textbf{BSD100} & \textbf{Urban100} & \textbf{Manga109} \\
 & & & &  & \textbf{Size} & \textbf{PSNR/SSIM} & \textbf{PSNR/SSIM} & \textbf{PSNR/SSIM} & \textbf{PSNR/SSIM} & \textbf{PSNR/SSIM} \\\hline
\multirow{3}{*}{Interp.}    & Nearest\cite{Bevilacqua2012} & CPU & 9ms & - & - & 26.25/0.737 & 24.65/0.653  & 25.03/0.629 & 22.17/0.615 & 23.45/0.741 \\
 & Bilinear\cite{Kirkland2010} & CPU & 20ms & - & - & 27.55/0.788 & 25.42/0.679  & 25.54/0.646 & 22.69/0.635 & 24.21/0.767 \\
 & Bicubic\cite{1981keys} & CPU & 97ms & - & - & 28.42/0.810 & 26.00/0.702  & 25.96/0.667 & 23.14/0.657 & 24.91/0.787 \\\hline
\multirow{2}{*}{SC} & ANR\cite{radu2013} & CPU & 1715ms & 1.43MB & - & 29.70/0.842 & 26.86/0.737 & 26.52/0.699  & 23.89/0.696 & 26.18/0.821 \\
   & A+\cite{radu2015} & CPU & 1748ms & 15.17MB & - & 30.27/0.860 & 27.30/0.750  & 26.73/0.709   & 24.33/0.719 & 26.91/0.848 \\\hline
\multirow{2}{*}{CNN} & CARN-M\cite{ahn2018} & CPU & 4955ms & 1.59MB & - & 31.82/0.890 & 28.29/0.775 & 27.42/0.731  & 25.62/0.769 & 29.85/0.899 \\
   & RRDB\cite{wang2019} & CPU & 31717ms & 63.83MB & - & \textbf{32.60}/\textbf{0.900} & \textbf{28.88}/\textbf{0.790}  & \textbf{27.76}/\textbf{0.743}   & \textbf{26.73}/\textbf{0.807} & \textbf{31.16}/\textbf{0.916} \\
   % & HF-CNN\cite{kim2018real} & FPGA & 16.6ms & 3.20KB & - & 36.51/0.9520 & 32.46/0.9055 & 31.27/0.8864  & 29.28/0.8916 & -/- \\
% & ERVSR\cite{sun2021fpga} & FPGA & 13.2ms & 3.81KB & - & 36.76/0.9553 & 32.51/0.9076 & 31.31/0.8887  & 29.30/0.8952 & 35.78/0.9682 \\
   \hline
 \multirow{10}{*}{LUT} & SRLUT\cite{Jo2021SRLUT}  & CPU & 94ms & 1.27MB & 15.24MB & 29.82/0.847 & 27.01/0.736 & 26.53/0.695 & 24.02/0.699 & 26.80/0.838  \\
  & SPLUT-S\cite{Ma2022SPLUT}  & CPU & 365ms & 5.5MB & 33MB & 30.01/0.852 & 27.20/0.743 & 26.68/0.702 & 24.13/0.706 & 27.00/0.843  \\
   & SPLUT-M\cite{Ma2022SPLUT} & CPU & 400ms & 7MB & 42MB & 30.23/0.857 & 27.32/0.746 & 26.74/0.704 & 24.21/0.709 & 27.20/0.848\\
   & SPLUT-L\cite{Ma2022SPLUT} & CPU & 822ms & 18MB & 108MB & 30.52/0.863 & 27.54/0.752 & 26.87/0.709 & 24.46/0.719 & 27.70/0.858 \\
 & MuLUT-SDY\cite{Li2022MuLUT} & CPU & 228ms & 3.82MB & 45.84MB & 30.40/0.860 & 27.48/0.751 & 26.79/0.709 & 24.31/0.714 & 27.52/0.855 \\
 & MuLUT-SDY-X2\cite{Li2022MuLUT} & CPU & 242ms & 4.06MB & 48.72MB & 30.60/0.865 & 27.60/0.754 & 26.86/0.711 & 24.46/0.719 & 27.90/0.863 \\
 & RCLUT\cite{Liu2023RCLUT} & CPU & 232ms & 1.51MB & 18.12MB & 30.72/0.868 & 27.67/0.758 & 26.95/0.715 & 24.57/0.725 & 28.05/0.865 \\
 & HKLUT-S\cite{huang2024HKLUT} & CPU & 66ms & \textbf{100KB} & 1.2MB & 30.35/0.859 & 27.39/0.748 & 26.73/0.706 & 24.23/0.711 & 27.38/0.852 \\%148.67
 & HKLUT-L\cite{huang2024HKLUT} & CPU & 99ms & 112.5KB & 1.35MB & 30.41/0.860 & 27.44/0.749 & 26.78/0.707 & 24.27/0.713 & 27.51/0.854 \\%223.01
 & Efficient LUT\cite{li2024} & FPGA & 3.08ms & 163KB & 652KB & 29.92/0.848 & 27.21/0.740 & 26.65/0.700 & 24.07/0.701 & 26.88/0.835 \\
  & Ours (RGB) & CPU/FPGA & 78/\textbf{2.35ms} & 133KB & 1.46MB & 30.42/0.861 & 27.44/0.750 & 26.78/0.708 & 24.28/0.713 & 27.51/0.855 \\%175.1
 & Ours (YUV) & CPU/FPGA & 74/\textbf{2.35ms} & 133KB & \textbf{532KB} & 30.25/0.856 & 27.34/0.746 & 26.75/0.705 & 24.22/0.710 & 27.08/0.842 \\%165.6
 \bottomrule
\end{tabular}}
}
\vspace{-0.2cm}
\label{tab:compare_4x_sr}
\end{table*}

\begin{table*}[!t]
\renewcommand{\arraystretch}{0.8}
\caption{Size and Quantitative comparison(PSNR/dB) for $4\times$ SR with varying stage, upscale factor, UV processing, kernel, and bit-width on 5 benchmark datasets.}
\vspace{-0.3cm}
\centering{
\setlength{\tabcolsep}{0.6mm}{
\begin{tabular}{lccccccccccc}
\toprule
\multirow{2}{*}{\textbf{Stage}} & \multirow{2}{*}{\textbf{Upscale}} & \textbf{UV} & \textbf{Kernel}& \textbf{Bit} & \multirow{2}{*}{\textbf{Size}}& \textbf{Actual} & \textbf{Set5}  & \textbf{Set14} & \textbf{BSD100} & \textbf{Urban100} & \textbf{Manga109} \\
 & & \textbf{Method} & \textbf{MSB/LSB} & \textbf{MSB/LSB} &  & \textbf{Size} & \textbf{PSNR/SSIM} & \textbf{PSNR/SSIM} & \textbf{PSNR/SSIM} & \textbf{PSNR/SSIM} & \textbf{PSNR/SSIM} \\\hline
 2 & $2\times2$  & Nearest & HDBL/HD & 4/4 & 133KB & 532KB & 30.26/0.857 & 27.35/0.746 & 26.75/0.706 & 24.22/0.710 & 27.08/0.843 \\
 2 & $1\times4$  & Nearest & HDBL/HD & 4/4 & 282KB & 1.128MB & 29.98/0.848 & 27.17/0.740 & 26.63/0.701 & 24.06/0.701 & 26.74/0.831  \\
  1 & $4$ & Nearest & HDBL/HD & 4/4 & 264KB & 1.056MB & 30.00/0.848 & 27.18/0.741 & 26.64/0.702 & 24.07/0.702 & 26.74/0.832 \\
 2 & $2\times2$ & Bicubic & HDBL/HD & 4/4 & 133KB & 532KB & 30.26/0.857 & 27.35/0.746 & 26.75/0.706 & 24.22/0.710 & 27.10/0.843 \\
 2 & $2\times2$  & Nearest & HDBL/HDB & 4/4 & 225KB & 900KB & 30.25/0.855 & 27.32/0.745 & 26.72/0.705 & 24.21/0.708 & 27.03/0.839 \\
  2 & $2\times2$  & Nearest & HDBL/HD & 3/5 & 33.5KB & 134KB & 30.06/0.851 & 27.20/0.741 & 26.65/0.701 & 24.09/0.704 & 26.75/0.834 \\
 \bottomrule
\end{tabular}}
}
\vspace{-0.2cm}
\label{tab:compare_4x_sr_ablation}
\end{table*}

\begin{table}[!t]
\renewcommand{\arraystretch}{0.8}
\caption{Performance summary of the proposed ScaleLUT and the hardware cost breakdown for the main computation blocks on ZCU102 WITH 300MHz}
\vspace{-0.3cm}
% \vspace{-0.1cm}
\centering{
\setlength{\tabcolsep}{0.6mm}
\begin{tabular}{cccccc}
\hline
\multirow{2}{*}{\textbf{Component}} & \multicolumn{4}{c}{\textbf{Resource Utilization}} & \multirow{2}{*}{\textbf{\begin{tabular}[c]{@{}c@{}}Power \\ (mW)\end{tabular}}} \\ \cline{2-5}
 & \textbf{LUT} & \textbf{FF} & \textbf{DSP} & \textbf{BRAM} &  \\ \hline
AXI-Regs & 369/334 & 812/662 & 0/0 & 0/0 & 11/6 \\
DMA & 2757/2753 & 349/363 & 0/0 & 0/0 & 30/93 \\
Padding and Ctrl. & 44/42 & 97/92 & 0/0 & 0/0 & 6/7 \\
Line buffer & 897/300 & 1050/348 & 0/0 & 7.5/2.5 & 40/8 \\
Pixel Arith. Unit & 7376/5007 & 5424/1808 & 0/0 & 0/0 & 85/18 \\
HDBL-LUT & 0/0 & 0/0 & 0/0 & 816/272 & 1909/660 \\
YUV-RGB Conv. & 0/2321 & 0/1036 & 0/0 & 0/0 & 0/130 \\
Others & 3210/748 & 900/984 & 0/0 & 0/0 & 17/79 \\
\textbf{Total (RGB)} & 19523 & 8632 & 0 & 823.5 & 2098 \\
\textbf{Total (YUV)} & 11505 & 5293 & 0 & 274.5 & 1001 \\ \hline
\end{tabular}
}
\vspace{-0.3cm}
\label{tab:FPGA_scalelut}
\end{table}

% \vspace{-0.5cm}
\section{Experiments}
\vspace{-0.1cm}
\subsection{Experiment Setup}
\vspace{-0.1cm}
\textbf{Datasets and Training:} We train the network model on the widely-used DIV2K training set\cite{Agustsson_2017_DIV2K} for SR task. Training runs for 200k iterations with a batch size of 16 on NVIDIA RTX 3090 GPUs using Adam optimizer ($\beta_1 = 0.9, \beta_2 = 0.999$, $\epsilon = 1e-8$) and MSE loss. Learning rate starts at $5 \times 10^{-4}$, decaying by 0.1 at 100k and 150k iterations. The input image is randomly cropped into $48\times48$ patches, and the dataset is enhanced by random rotation and flipping.

\textbf{Evaluation:} We evaluate on standard benchmarks including Set5\cite{Bevilacqua2012}, Set14\cite{Zeyde2012}, BSD100\cite{Martin2001}, Urban100\cite{Huang-CVPR-2015}, Manga109\cite{Matsui2015} for SR. We compare against SOTA methods including classical Interpolation methods (Nearest\cite{Bevilacqua2012}, Bilinear\cite{Kirkland2010}, Bicubic\cite{1981keys}), classical sparse-coding methods (ANR\cite{radu2013}, A+\cite{radu2015}), CNN-based methods (CARN-M\cite{ahn2018}, RRDB\cite{wang2019}), and LUT-based approaches (SR-LUT\cite{Jo2021SRLUT}, SPLUT-S, SPLUT-M and SPLUT-L\cite{Ma2022SPLUT}, MuLUT-SDY and MuLUT-SDY-X2\cite{Li2022MuLUT}, RCLUT\cite{Liu2023RCLUT}, HKLUT-S and HKLUT-L\cite{huang2024HKLUT}, Energy-efficient LUT\cite{li2024}). We use Peak Signal-to-Noise Ratio (PSNR) and Structural Similarity Index (SSIM) as evaluation metrics. 

\textbf{Hardware system setup:} We use an AMD Zynq UltraScale+ ZCU102 FPGA with an ARM Cortex-A53 CPU and 4 GB DDR4 as the main experimental platform. A Zidoo Z9X PRO video player (480p–4K) feeds input through the HDMI RX port, while the HDMI TX port outputs the processed video to a 4K display. This setup enables real-time validation of our LUT-based video processing system and provides a comprehensive assessment of its performance.

\begin{table*}[!t]
\renewcommand{\arraystretch}{0.90}
\caption{Performance comparison of ScaleLUT with state-of-the-art SR accelerators. All designs are evaluated on 2$\times$ FHD-to-4K upscaling, reporting FPGA resource usage, throughput, and energy efficiency. ScaleLUT additionally supports arbitrary power-of-two scales (2$^n\times$).}

% \vspace{-0.3cm}
\centering{
\setlength{\tabcolsep}{0.65mm}
\begin{tabular}{ccccccccccccc}
\hline
\multirow{2}{*}{\textbf{Design}} & \multirow{2}{*}{\textbf{Category}} & \multirow{2}{*}{\textbf{\begin{tabular}[c]{@{}c@{}}FPGA \\ Platform\end{tabular}}} &\multirow{2}{*}{\textbf{\begin{tabular}[c]{@{}c@{}}Supported \\ Scale\end{tabular}}} & \multirow{2}{*}{\textbf{\begin{tabular}[c]{@{}c@{}}Fmax \\ (MHz)\end{tabular}}} & \multirow{2}{*}{\textbf{\begin{tabular}[c]{@{}c@{}}Norm.\\FPS\end{tabular}}} & \multicolumn{4}{c}{\textbf{Resource Utilization}} & \multirow{2}{*}{\textbf{\begin{tabular}[c]{@{}c@{}}Power \\ (mW)\end{tabular}}} & \multirow{2}{*}{\textbf{\begin{tabular}[c]{@{}c@{}}Throughput\\ (Mpixels/s)\end{tabular}}} & \multirow{2}{*}{\textbf{\begin{tabular}[c]{@{}c@{}}Energy Eff.\\ (Mpixels/J)\end{tabular}}} \\ \cline{7-10}
  &  &  &  &  &  & \textbf{LUT} & \textbf{FF} & \textbf{BRAM\_36K} & \textbf{DSP} &  &  &  \\ \hline
\cite{taito2016}-2016 & CNN       & XCVU095    & 2$\times$ & 133  & 12    & 266K  & 15K   & 131   & 768  &  -    & 94.9 & - \\
\cite{he2018fpga}-2018      & CNN       & ZC706     & 2$\times$ & 100  & 31.7     & 104K  & 66K   & 163   & 858  &  -    & 250.8 & - \\
\cite{kim2018real}-2019     & CNN       & XCKU040   & 2$\times$ & 150  & 76.0     & 151K  & 121K  & 43.5  & 1920 & 5686 & 600.0 & 105.5 \\
\cite{chang2020}-2020 & CNN       & XC7K410T    & 2/3/4$\times$ & 130  & 62.7    & 167K  & 158K   & 205   & 1512  &  5380   & 495.7 & 92.1 \\
\cite{sun2021fpga}-2022     & CNN       & XCKU15P   & 2$\times$ & 160  & 80.5     & 98K   & 57K   & 148   & 1820 & 5470 & 637.0 & 111.0 \\
\cite{li2024}-2024 & LUT                & XCKU040   & 2$\times$ & 150  & 75.5     & 29K   &  9K  &  \textbf{225}   & 12  &  1720         & 597.2  & 347.2 \\
Ours (RGB) & LUT                        & ZCU102    & \textbf{2$^n$$\times$} & \textbf{300} & \textbf{95.3} & 20K & 8.6K & 823.5 & \textbf{0} & 2098 & 753.4 & 359.2 \\
Ours (YUV) & LUT                        & ZCU102    & \textbf{2$^n$$\times$} & \textbf{300} & \textbf{95.3} & \textbf{12K} & \textbf{5.3K} & 274.5 & \textbf{0} & \textbf{1001} & \textbf{753.6} & \textbf{752.8} \\ \hline
\end{tabular}}
% \vspace{-0.3cm}
\label{tab:resource_com}
\end{table*}

% \vspace{-0.1cm}
\subsection{Algorithm Performance Comparison}
\vspace{-0.1cm}

Table~\ref{tab:compare_4x_sr} compares the proposed YUV and RGB ScaleLUT variants with representative SR methods. The RGB variant applies the proposed LUT architecture to all R/G/B channels, while the YUV variant performs LUT-based SR only on the Y channel and handles U/V separately. Therefore, the YUV design requires only one-third of the multi-channel LUT storage used by RGB-domain deployment. Here, actual size denotes the physical memory footprint after deploying multi-channel and multi-angle rotation kernels on hardware.

The proposed YUV design achieves the smallest hardware memory footprint among LUT-based methods while maintaining competitive PSNR/SSIM. It improves over Efficient-LUT by about 0.3 dB and achieves comparable quality to RGB-based LUT methods with much larger deployed sizes. The RGB variant further outperforms similarly sized HKLUT-S and larger HKLUT-L, demonstrating the effectiveness of the proposed two-stage LUT design. Overall, ScaleLUT provides a better quality--memory trade-off than prior LUT-based SR methods.

Table~\ref{tab:compare_4x_sr} also reports runtime for $4\times$ SR with a $1280\times720$ output. Benefiting from the fully parallel FPGA implementation, ScaleLUT achieves the shortest runtime among all compared methods, with at least $28\times$ speedup over the fastest CPU-based HKLUT result. The CPU runtime of our method is also reported for reference, confirming that the FPGA acceleration preserves the same algorithmic output while substantially improving inference speed.

\subsection{Ablation Study}

To validate the proposed hardware-oriented design framework, Table~\ref{tab:compare_4x_sr_ablation} reports design points from the explored search space, covering stage configuration, upscale strategy, UV processing, kernel design, and bit allocation. Each design is compared in terms of model size, actual deploy size after rotation replication, and PSNR/SSIM on five benchmark datasets. Other combinations are omitted because they either incur higher hardware cost with negligible quality gain or show inferior quality--efficiency trade-offs.

The ablation results show three key observations. First, progressive two-stage upscaling ($2\times2$) reduces model size by nearly 50\% compared with direct $1\times4$ upscaling while improving PSNR by at least 0.12 dB, indicating that stage-wise reconstruction better balances quality and storage cost. Second, more complex UV processing brings limited benefit: replacing nearest-neighbor UV processing with bicubic interpolation only marginally improves PSNR, e.g., 0.02 dB on Manga109, but introduces additional multiplier overhead. Third, kernel design and bit allocation strongly affect the quality--efficiency trade-off. Adding more LSB kernels increases deploy size without PSNR improvement, while changing the MSB/LSB allocation from 4/4 to 3/5 greatly reduces model size but consistently degrades PSNR.

These results confirm that the final ScaleLUT architecture is selected through systematic design-space exploration under joint consideration of reconstruction quality and hardware efficiency, rather than heuristic manual design.

\subsection{System Performance Comparison of Hardware Metrics}

As summarized in Table~\ref{tab:FPGA_scalelut}, the RGB-format ScaleLUT implementation on ZCU102 uses 20K LUTs, 8.6K FFs, 274.5 BRAMs, and zero DSPs, with 2098 mW on-chip power. In contrast, the YUV implementation uses only 12K LUTs and 5.3K FFs with the same BRAM usage and zero DSPs, reducing LUT and FF consumption by 40.0\% and 38.4\%, respectively, while lowering power to 1001 mW. This reduction is achieved with only about 0.1 dB PSNR loss, demonstrating the practicality of the proposed YUV-domain design.

Table~\ref{tab:resource_com} compares ScaleLUT with SOTA FPGA-based SR accelerators, including CNN-based designs~\cite{taito2016,he2018fpga,kim2018real,chang2020,sun2021fpga} and LUT-based SR~\cite{li2024}. ScaleLUT operates at 300 MHz, nearly twice the frequency of most prior designs, benefiting from its multiplier-free LUT architecture and balanced pipeline. Unlike CNN accelerators with MAC-intensive datapaths or prior LUT designs with interpolation-related multipliers/dividers, ScaleLUT uses shift-based arithmetic and fully pipelined parallel processing.

Consequently, ScaleLUT achieves clear advantages in resource, throughput, and energy efficiency. Compared with the fastest CNN-based design~\cite{sun2021fpga}, it improves throughput by 18.4\% and energy efficiency by 6.76$\times$. Compared with the prior LUT-based accelerator~\cite{li2024}, it reduces LUT and FF usage by 58.6\% and 41.1\%, respectively, while improving throughput by 26.2\% and energy efficiency by 2.17$\times$. In addition, the optimized dataflow and configurable control logic support arbitrary $2^n$ upscaling factors, enabling flexible deployment across practical SR scenarios.
\section{Conclusion}

\begingroup
\emergencystretch=.8em
ScaleLUT effectively addresses the long-standing challenge of achieving real-time, high-quality multi-scale SR on resource-constrained edge devices. By adopting a unified hardware-oriented LUT design framework, ScaleLUT introduces a fully parallel and configurable LUT-based framework that fully exploits the multiplication-free nature of LUT inference, thereby eliminating MAC-intensive computation. With a hardware-efficient YUV-domain design and rotation-ensemble RF coverage, ScaleLUT substantially reduces LUT complexity and storage overhead without sacrificing reconstruction fidelity. Implemented on a Xilinx ZCU102 FPGA, ScaleLUT achieves 95.3 FPS at 300 MHz for 4K $\times 2$ SR, while reducing LUT usage by 58.6\%, flip-flops by 41.1\%, and power consumption by 42\%, all with zero DSP utilization. These results demonstrate that ScaleLUT provides a simple, scalable, and highly efficient solution for real-time SR, highlighting the strong potential of LUT-based architectures as practical lightweight alternatives to power-hungry deep-learning SR accelerators for diverse edge deployment scenarios.
\par
\endgroup

% \printbibliography


\clearpage
%%%%%%%%%%%%%%%%%%%%%%%%%%%%%%%%%%%%%%%%%%%%%%%%%%%%%%%%%%%%%%%%%%%%%%%%%%%%%%%%%%%%%%%%%%%%%%%%%%%%%%%%%%%%%%%%%%
% \section*{Acknowledgement}
% \label{sec:conclusion}
% This work was supported in part by the Theme-based Research Scheme (TRS) project T45-701/22-R, National Natural Science Foundation of China (62404187) and the General Research Fund (GRF) Project 17203224, of the Research Grants Council (RGC), Hong Kong SAR.
\bibliographystyle{unsrt}
\bibliography{references}

\end{document}